\documentstyle[aps,prb,epsf,floats]{revtex}

\def\slz{{\sl SL\/}(2,{\bf Z})}
\def\xy{{\it x-y\/}}

\def\Incr{{\setbox0\hbox{$\triangle$}\hbox%
    to\wd0{\hfill\hbox{$\cdot$}\hfill}\hskip-\wd0\box0}} 
\makeatletter\@addtoreset{paragraph}{section}\makeatother

\begin{document}
\twocolumn[\hsize\textwidth\columnwidth\hsize\csname @twocolumnfalse\endcsname

\title{Global symmetries of quantum Hall systems: lattice description} 
%% Lattice version of the CSLG model: exact results
\author{Leonid P. Pryadko} 
\date{February 24, 1997} % final version for submission
%\date\today
\maketitle

\begin{abstract}
  I analyze non-local symmetries of finite-size Euclidean 3D lattice
  Chern-Simons models in the presence of an external magnetic field
  and non-zero average current.  It is shown that under very general
  assumptions the particle-vortex duality interchanges the total
  Euclidean magnetic flux ${\bf \Phi}/2\pi$ and the total current
  ${\bf I}$ in a given direction, while the flux attachment
  transformation increases the flux in a given direction by the
  corresponding component of the current, ${\bf
    \Phi}/2\pi\rightarrow{\bf \Phi}/2\pi+2{\bf I}$, independently of
  the disorder.  In the language of $2\!+\!1$ dimensional models,
  appropriate for describing quantum Hall systems, these
  transformations are equivalent to the symmetries of the phase
  diagram known as the Correspondence Laws, and the 
  non-linear current--voltage mapping between mutually dual points,
  recently observed near the quantum Hall liquid--insulator
  transitions.
\end{abstract}
\vskip2pc]
\pagestyle{myheadings}
\markright{{\rm ``Global symmetries in quantum Hall systems: lattice
    description''} by Leonid Pryadko} 
\narrowtext
\section*{Introduction}

Beginning with the analysis of the Ising model by Kramers and
Wannier\cite{Kramers-41} in early forties, the particle--vortex
duality found extensive use in physics.  However, duality had been
always considered as a kind of ``theoretical'' symmetry, whose
manifestations are limited to the critical region and generally are
very hard to observe.  This is one of the reasons why the
experiments\cite{Shahar-95B,Shahar-96A,Shahar-96B} observing a
precise duality between the adjacent quantum Hall phases came as a
total surprise.  The agreement with the apparently simple
theory\cite{Kivelson-92} combining the duality transformation and the
random phase approximation (RPA) went far beyond the original
expectations of the authors.  In fact, experimentally\cite{Shahar-95B}
the duality was shown to relate the entire set of non-linear $I$--$V$
curves in a surprisingly wide region of temperatures, gate voltages
and magnetic fields, indicating that the dynamics on the two sides
of the transitions must be very similar.  This is in spite of the fact
that theoretically, according to the very same duality mapping, the
elementary particle excitations at the two sides of the transition,
fermions and fractionally charged vortex-like quasiparticles, are very
different!  Even more surprising are the experimental indications of
duality between the two phases in the metal--insulator phase
transitions\cite{Simonian-96,Kravchenko-96}, apparently observed in
silicon MOSFETs in the absence of a magnetic field, where the
conventional theory of localization predicted no phase transition at
all.

The fractionally charged quasiparticles were introduced by
Laughlin\cite{Laughlin-83A} as excitations on top of the
highly-correlated state now known as the Laughlin liquid; these
objects were used to explain the observed fractional values of the
Hall conductance.  Unfortunately, the wavefunction description does
not easily reveal the order parameter associated with different phases
and it could not be used to study the phase transitions between the
plateaus.  The interpretation of the peculiar off-diagonal long-range
order inherent to Laughlin's wavefunction as the condensation of the
composite particles was proposed\cite{Girvin-87} by Girvin and
MacDonald.  These ideas were later developed\cite{Zhang-89} by Zhang,
Hansson and Kivelson, who derived the Chern-Simons-Landau-Ginzburg
(CSLG) theory\cite{Zhang-92} directly from the microscopic
Hamiltonian.  The symmetries of the ground-state wavefunction now
became apparent as the (approximate) non-trivial symmetries of the
model itself.  These symmetries impose many important constraints on
the allowed phase transitions and were used\cite{Kivelson-92} by
Kivelson, Lee and Zhang (KLZ) to construct the Global Phase Diagram of
the quantum Hall system.

Despite all the success of the CSLG % Chern-Simons-Ginzburg-Landau 
theory and its excellent experimental confirmation\cite{Wei-88,%
  Engel-90,Koch-91A,Glozman-95,Shahar-95A,Shahar-95B,Shahar-96A,%
  Shahar-96B}, some questions still remain unanswered.  The most
important of these questions is why does so simple a model work so
well.  The original KLZ calculations were done only within the RPA
approximation, and the universality of the scalar field correlators
was conjectured rather than derived.  Indeed, the quasiparticles have
different charges and even their fractional statistics differ, and one
would think that the global universality in this picture is not likely
to happen.  For example, calculations\cite{Wen-Wu-93,Wei-93},
perturbative in the Chern-Simons coupling, indicate that this
additional gauge coupling is at least a marginally relevant
perturbation to the model of free bosons.  Similar
calculation\cite{Pryadko-94}, performed without the $1/N$ expansion
suppressing the effect of gauge degrees of freedom, demonstrated that
the Chern-Simons gauge interaction is actually a {\em relevant\/}
perturbation to the scalar field fixed point; this interaction leads
to a first-order phase transition in the regime where perturbation
theory applies.

A formal symmetry-based approach to the problem of the quantum Hall
phase transitions was initiated\cite{Lutken-92,Lutken-93} by Lutken
and Ross, and later further developed\cite{Damgaard-96,Burgess-96} to
apply for the beta-function flows.  The idea was to find all the
consequences of the observed \slz\ symmetry group without specifying
how this symmetry is achieved.  In the spirit of the two-parameter
scaling model of the quantum Hall
effect\cite{Khmelnitskii-83,Pruisken-localization}, it was postulated
that the partition function and the scaling beta-function for the
complexified conductance $\Sigma=\Sigma_{xy}+i\Sigma_{xx}$ of the
system, are invariant under the elements of \slz\ and can depend on
$\Sigma$ only as ``quasi-holomorphic'' functions.  This assumption,
together with the known behavior in the limits of weak and strong
coupling, severely restricts the phase diagram and the possible
functional form of beta-function.  Unfortunately, so far nobody
has managed to complete this program and find the critical exponents in
the vicinity of the critical point relying on the symmetry
considerations, alone.

As an alternative to the conventional perturbative approach,
S.-C.~Zhang and the author attempted\cite{Pryadko-95} to understand
the universality of the quantum Hall transitions using the lattice
\xy\ model coupled with the Chern-Simons gauge field, which comes
naturally as the strong-coupling regime of the CSLG model.  They
constructed the exact duality and flux attachment transformations on
the lattice and proved that the Chern-Simons field is the most
relevant interaction in this problem and, modulo the irrelevant
higher-order in momenta terms, the phase transitions in different
Chern-Simons models with the coefficients related by these
transformations must be in the same universality class.  Fradkin and
Kivelson\cite{Fradkin-96}, in addition to the model similar to that of
Ref.\CITE{Pryadko-95}, considered a model where the gauge-field
polarization operator acquires an additional $T$-symmetric, linear in
momenta part.  The corresponding dimensionless coupling constant,
combined with the usual Chern-Simons coefficient, forms a
complex-valued parameter $z$ which transforms according to a subgroup
of the modular group \slz.  In particular, there are infinitely many
fixed points that remain invariant under certain elements of this
group.  It was proposed that corresponding values of the parameter $z$
must coincide with the critical points of the quantum Hall system.
Once again, symmetry requires the conductivity in these critical
points to have certain (generally different for different fixed
points) universal values.

The major advantage of Chern-Simons \xy\ models is that they allow to
treat the gauge fields, apparently the most relevant part of the
interaction, non-perturbatively, considering all other interactions as
less relevant.  On the other hand, the lattice models considered so
far did not have precisely the same symmetries as the continuum CSLG
model derived\cite{Zhang-89} directly from the microscopic Hamiltonian
of electrons.  Specifically, the lattice
models\cite{Pryadko-95,Fradkin-96} have the exact rotational symmetry
(or relativistic one after the Wick rotation) while the scalar field
in Refs.\CITE{Zhang-89,Kivelson-92} is non-relativistic.  In addition,
none of the lattice models was constructed carefully enough to allow
the finite-size scaling analysis in the presence of external magnetic
field, current or charge density.

The goal of this work is to construct a lattice Chern-Simons model as
a regularization of the continuum CSLG model in the limit of strong
short-range electron-electron repulsion, at the same time keeping
track of the long-distance properties of the model.  I show that the
required model is exactly the Chern-Simons \xy\ model, extended to
include a non-zero flux of external magnetic field and a non-zero
average current density.  Even though this model is written in
rotationally (relativistically) {\em covariant\/} form, the Lorentz
symmetry is broken by the external current (the last component of
which corresponds to the charge density) and the external magnetic
field, so that both by construction and by its symmetry the model
represents a quantum Hall system.  It is shown that under very general
assumptions, in addition to transformations of the Chern-Simons
coupling\cite{Pryadko-95}, the duality interchanges the total magnetic
flux ${\bf \Phi}/2\pi$ and the total current ${\bf I}$ in a given
direction, while the flux attachment transformation increases the
total magnetic flux in a given direction by the corresponding
component of the total electrical current, ${\bf
  \Phi}/2\pi\rightarrow{\bf \Phi}/2\pi+2{\bf I}$.  

The derived transformations, after being reinterpreted in terms of the
$2\!+\!1$ dimensional systems, are equivalent to the symmetries of the
phase diagram known as Correspondence Laws\cite{Kivelson-92}, as well
as the
nonlinear current--voltage mapping between mutually dual points,
recently observed\cite{Shahar-95B} near the quantum Hall
liquid--insulator transitions.  Even though the exact symmetries of
the phase diagram can be derived within other approaches, like, for
example, the 2D CFT formulation\cite{Cristofano-97}, only within the
CS \xy\ model one can understand the non-linear transport
relationships as well.  Unlike the previously studied continuum models
where the problems of renormalization are unavoidable, here the
current--voltage mapping is {\em exact\/} and is an immediate
consequence of the duality regardless of the specific form of the
gauge and scalar-field coupling.  Moreover, these symmetries hold in
the presence of arbitrary potential and magnetic disorder, which
remain unchanged by either the duality or the flux attachment (up to a
short-radius formfactor coming from the different size of the
quasiparticles in different representations).  This implies that the
derived results will hold even if the disorder, marginal by the power
counting, turns out to be a relevant perturbation.

The Section~\ref{sec:finite-system} introduces the lattice
Chern-Simons \xy\ model in an external gauge field as the lattice
regularization of the phase part of the CSLG model after integration
over the density fluctuations.  The infrared behavior of this model is
controlled by defining it on a torus using the magnetic translation
group.  The duality and flux attachment transformations of the lattice
model are discussed in Sections~\ref{sec:3d-duality}
and~\ref{sec:fa-transform} respectively.  The application of derived
results to the real-time quantum Hall systems, including the
derivation of the Correspondence Laws\cite{Kivelson-92} and the
current--voltage duality, are presented in
Sec.~\ref{sec:correspondence}.

Before ending this Introduction, I would like to remark that there
have been a great number of publications on the problem of phase
transitions and the duality in the quantum Hall system, and it would
be impossible to list even all the approaches here.  An excellent
summary of plausible interpretations of the
experiment\cite{Shahar-95B} and the key difficulties associated with
these interpretations was recently given\cite{Shimshoni-96} by
Shimshoni, Sondhi and Shahar.

\section{Definition of the model}
\label{sec:finite-system}

\subsection{Lattice and continuum Chern-Simons models}

The Euclidean partition function of the Chern-Simons \xy\ model with
an external field can be written as
\begin{equation}
  \label{eq:def-csxy}
  {\cal Z}= \prod\int d{\bf a}\,d\theta\, %
  e^{-i{\cal S}_\theta(\theta,{\bf a}+{\bf A})-i{\cal S}_a({\bf a})} ,
\end{equation}
where the gauge part
\begin{equation}
  \label{eq:def-gauge-part}
  i{\cal S}_a={1\over2}\int {d^3k \over (2\pi)^3} %
  {\bf a}_{-{\bf k}}{\cal K}({\bf k}){\bf a}_{\bf k}, 
\end{equation}
and scalar part
\begin{equation}
  \label{eq:def-scalar}
    i{\cal S}_\theta=-{1\over T}\sum\cos(\nabla\theta-{\bf a}-{\bf A})
\end{equation}
of the action depend on the phases $\theta$ defined at every vertex of
the three-dimensional cubic lattice, while the Chern-Simons gauge
field ${\bf a}$ and external gauge field ${\bf A}$ live on the links.
As usual for lattice gauge theories, the value of ${\bf A}$ on the
link can be interpreted as the integral of the physical gauge field
along this link
\begin{displaymath}
  {\bf A}_{n,n+\hat\mu}=\int_{{\bf r}_n}^{{\bf r}_{n+\hat\mu}}
  {\bf A}_{\mu}({\bf x}) d{\bf x}^\mu;
\end{displaymath}
for notational convenience, in this paper the contour sums of gauge
quantities are denoted as the continuum integrals along the
appropriate contours.  Similarly, the flux of the magnetic field
through some area, a sum over plaquettes in the discrete lattice
model, will be denoted as the continuum surface integral of the field
over this area.  These notations were chosen for the sake of
readability of the paper.  The Reader interested in the details of the
lattice definition of the Chern-Simons term is directed to
Ref.~\CITE{Pryadko-95}.

The dynamics of the field ${\bf a}$ is determined by the transverse
gauge kernel ${\cal K}({\bf k})$.  The most general
rotationally-invariant form of this kernel can be written as the sum
of $T$-antisymmetric and $T$-symmetric parts
\begin{equation}
  \label{eq:gen-cs-kernel}
  {\cal K}^{\mu\nu}({\bf k})=
  f(k)e^{\mu\nu\rho}{\bf P}_\rho+
  g(k)\left(P^2\delta^{\mu\nu}-{\bf P}_\mu{\bf P}_\nu\right),
\end{equation}
depending on the lattice momentum ${\bf P}_\mu=2\sin(k_\mu/2)$.  It is
the small-momentum behavior of the functions $f$ and $g$ that
determines the properties of the model, as has been shown in
Refs.\CITE{Pryadko-95,Fradkin-96}.  Specifically, the usual
Chern-Simons model can be defined by the condition
\begin{equation}
  \label{eq:def-purely-cs}
  \lim\limits_{k\rightarrow0}f(k)={f_0\over2\pi}
  ={1\over2\pi\,\alpha}\quad {\rm
    and}\quad \lim\limits_{k\rightarrow0}k g(k)=0.
\end{equation}
The generalization of this standard form considered by Fradkin and
Kivelson\cite{Fradkin-96}, has both limits finite,
\begin{equation}
  \label{eq:def-genrl-cs}
  \lim\limits_{k\rightarrow0}f(k)={f_0\over2\pi}\quad {\rm and}\quad
  \lim\limits_{k\rightarrow0}k g(k)={g_0\over2\pi}.
\end{equation}
Unlike the usual Chern-Simons model, the model with such a kernel has
self-dual fixed points; they were associated\cite{Fradkin-96} with the
fixed points of the quantum Hall phase transitions.  

As defined, the partition function~(\ref{eq:def-csxy}) is symmetric
with respect to simultaneous change of the sign of all components of
the fields $\nabla\theta$, ${\bf a}$ and ${\bf A}$; since the phase
gradient is associated with the current, both the external magnetic
field and the current are reversed,
\begin{equation}
  \label{eq:inversion}
  {\bf j}\rightarrow-{\bf j},\quad
  {\bf a}\rightarrow-{\bf a},\quad
  {\bf A}\rightarrow-{\bf A}.
\end{equation}
This trivial symmetry preserves the handedness of the coordinate
system, and therefore does not change the sign of the
pseudoscalar~$\alpha$.  One can also change the handedness of the
coordinate system, simultaneously reversing the sign of $\alpha$ and
all pseudovectors like the magnetic field.  In terms of the global
flux of the total magnetic field ${\bf \Phi}=\int
\nabla\!\times\!\left({\bf A}+{\bf a}\right) \,d{\bf s}$, this
transformation can be written as
\begin{equation}
  \label{eq:reflection}
  {\bf \Phi}=-{\bf \Phi}',
\end{equation}
while polar vectors, (for example, the particle current density ${\bf
  j}$), remain unchanged.

At first glance, the lattice model~(\ref{eq:def-csxy}) seems to be
only vaguely related to the quantum Hall problem.  As defined, this
model treats all coordinates evenly, unlike the continuum CSLG
model\cite{Zhang-89,Zhang-92} with the action
\begin{equation}
  \label{eq:def-cslg}
  S= S_a+ S_{\varphi}
   =\int dtd^2x\,{\cal L}_a+\int dtd^2x\,{\cal L}_{\varphi},
\end{equation}
which has a similar gauge part 
$ %
  {\cal L}_a={e^{\mu\nu\rho}a_\mu\partial_\nu a_\rho}/ ( 4\pi\alpha)
$, %
but an explicitly non-relativistic scalar part
\begin{eqnarray}
  \label{eq:def-cslg-scalar}
  {\cal L}_{\varphi}&=& \varphi^{\dag}\left(i\partial_t-a_t
    -{A_t}\right)\varphi \\ &\relax & %
  -{1\over2m_e}\left|\left(i\vec\nabla+\vec{a}+\vec{A}\right)\varphi\right|^2 
  -{V\over2} (\varphi^{\dag}\varphi-\overline\rho)^2\nonumber
\end{eqnarray}
defined in terms of the wavefunction
$\varphi=\sqrt{\rho}\exp{i\theta}$ of the bosonic condensate, where
$\theta$ and $\rho$ are the corresponding microscopic phase and 
density.  

Note, however, that even in the absence of the gauge fields, the
lattice model~(\ref{eq:def-csxy}) represents only the phase sector of
the continuum model~(\ref{eq:def-cslg}); any such model should have a
linear-spectrum mode as long as there is a finite phase stiffness
generating the propagation velocity $v$.  Formally, the
relativistically-symmetric form of the phase action can be
obtained\cite{Fradkin-book}
%%% other reference??? [...]
by integrating away the fluctuations of the density
$\rho=\varphi^{\dagger}\varphi$ near its average value
$\overline\rho$, which leads to the Lagrangian
\begin{equation}
  \label{eq:density-integrated}
  {\cal L}_{\theta}={k\over2}\left[
      {1\over v^2}(\partial_t\theta+a_t+A_t)^2 
      -\left(\vec\nabla\theta+\vec{a} +\vec{A}\right)^2\right]
\end{equation}
dependent on the phase $\theta$ only.  Here $k=\overline\rho/m_e$ and
$v^2=\overline\rho V/m_e$ are the effective stiffness and velocity of
the phase excitations calculated with frozen gauge fields; certainly,
these values should be understood as bare values of the parameters.

Even after the density integration, the ``relativistic'' symmetry of
the model~(\ref{eq:density-integrated}) is not exact: it is broken by
the external gauge field ${\bf A}$ which reflects the presence of a
uniform magnetic field and both scalar and magnetic disorder.  After
the Wick rotation $t\rightarrow i\tau$, $A_t\rightarrow -iA_z$, the
normal component of the magnetic field 
$ %
  B=\partial_x {\bf A}_y-\partial_y {\bf A}_x
$ %
becomes the last component of the Euclidean magnetic field $B_z=B$,
while the components of the lateral electric field 
\begin{displaymath}
  \vec{E}=-\vec\nabla A_t-\partial_t \vec{\bf A}
\end{displaymath}
correspond to the remaining components 
$ %
  B^j=i\varepsilon^{ji} E_i;
$ %
clearly, the analytic continuation from the imaginary values of
external fields is needed to relate the results of the two approaches.
Similarly, physical charge density $\rho$ and the current density
$\vec{\hbox{\it\j}}$ determine the three components of the Euclidean
current density ${\bf j}^\mu$ whose average values can be obtained
from the partition function~(\ref{eq:def-csxy}) by differentiating
with respect to the components of the Euclidean gauge field,
\begin{equation}
  \label{eq:euclidean-current}
  \langle{\bf j}^\mu\rangle=-{1\over i}{\delta\log Z\over\delta A_\mu}.
\end{equation}
This implies the correspondence ${\bf
  j}^\mu\rightarrow(\rho,\,i\vec{\hbox{\it\j}})$; the spatial
components of the Euclidean current acquire imaginary value and, once
again, the analytic continuation must accompany the Wick rotation.
 
\subsection{Magnetic translations and periodic boundary conditions}
By going from the continuum Lagrangian~(\ref{eq:density-integrated})
to the lattice action~(\ref{eq:def-scalar}) we regularized all
possible ultra-violet divergences.  To finish the definition of the
model, it is now necessary to regularize its infrared behavior.  This
can be done by defining the model on a closed surface like a torus.
Let us imagine the finite-size system of interest as a unit cell of an
infinite lattice, where all gauge-invariant quantities are perfectly
periodic.  On the other hand, a gauge field {\em cannot\/} obey the
periodic boundary conditions in any two directions if it has a
non-zero magnetic flux in the perpendicular direction.  The exact
periodicity in such a system can be established using the operators of
magnetic translation, defined as the combination of the ordinary
translation and the special gauge transformation restoring the
original form of the Hamiltonian.

As an example, consider a simplified version of lattice
model~(\ref{eq:def-scalar}) where the Chern-Simons field has been
frozen,
\begin{equation}
  \label{min-coupling-xy}
  i{\cal S}_{\rm min}= %
  -{1\over T}\sum_{\bf r} \cos\left(\nabla \theta-{\bf A}\right),
\end{equation}
which is just the three-dimensional \xy\ model minimally coupled to an
external magnetic field.  Even though the vector potential changes
under the gauge transformation
\begin{equation}
  \theta\rightarrow\theta+f,\quad %
  {\bf A}\rightarrow {\bf A}+\nabla f,
  \label{gauge-transf}
\end{equation}
the magnetic field ${\bf B}=\nabla\!\times\!{\bf A}$ and all physical
observables remain invariant.

Similarly, a displacement by a period ${\bf L}_\mu$ in a given
direction $\mu=x$, $y$ or $z$ changes the vector potential by the
increment defined as $\Incr_\mu{\bf A}({\bf r}) \equiv{\bf A}({\bf
  r}+{\bf L}_\mu)-{\bf A}({\bf r})$.   Since the
gauge-invariant magnetic field $\nabla\!\times\!{\bf A}$ is periodic
by assumption, its increment $\Incr_\mu$ vanishes,
$\Incr_\mu(\nabla\!\times\!{\bf A})=\nabla\!\times\!\Incr_\mu{\bf
  A}=0$, so that the increment $\Incr_\mu$ of the vector potential
${\bf A}$ can be expressed as a gradient,
\begin{equation}
  \Incr_\mu{\bf A}({\bf r})=\nabla F_\mu({\bf r}).
  \label{patching-f-defined}
\end{equation}
The newly-defined scalar function generates the gauge
transformation~(\ref{gauge-transf}) with $f=-F_\mu$, which exactly
restores the original form of the model.  Therefore, even though the
exact form of the model~(\ref{min-coupling-xy}) is changed by both the
translations and gauge transformations, their combination preserves
each term of this model.  Such combination of a translation and a
gauge transformation define the ``magnetic translation'' $\hat T({\bf
  L}_\mu)$.  As in the case of uniform magnetic field, the magnetic
translations $\hat T({\bf L}_\mu)$ and $\hat T({\bf L}_\nu)$ commute
with each other only if the magnetic flux through the area ${\bf
  L}_\mu\!\times\!{\bf L}_\nu$ is equal to an integer number of flux
quanta $\phi_0=h c/e$.

Let us utilize the magnetic translation symmetry to specify the
periodic boundary conditions for our lattice system in the region
$0\leq {\bf r}_i<{\bf L}_i$.  The angle $\theta$ in the point
$(x,y,{\bf L}_z)$ outside the boundaries of this region must be
replaced by the appropriately corrected angle in $(x,y,0)$ using the
symmetry with respect to magnetic translations
\begin{equation}\label{patching-cond}
  \theta(x,y,{\bf L}_z)- F_z(x,y,{\bf L}_z)\equiv\theta(x,y,0);
\end{equation}
similar patching conditions must be defined for the two remaining
``outer'' surfaces of the system.  The
expression~(\ref{patching-cond}) contains only the part of the
function $F_z(x,y,{\bf L}_z)$ at the ``top'' face $z={\bf L}_z$ of the this region
of interest; similarly only the values $F_x({\bf L}_x,y,z)$ and
$F_y(x,{\bf L}_y,z)$ of two other patching functions at their corresponding
``upper'' faces are relevant.  Each of these in-plane values obeys a simpler
two-dimensional equation
\begin{equation}
  \label{patching-f-in-twod}
  \Incr_\mu {\bf A}_\perp=\nabla_\perp F_\mu
\end{equation}
instead of Eq.~(\ref{patching-f-defined}), where the index ``$\perp$''
denotes the part of the vector orthogonal to the axis ``$\mu$''.  In
the following it is assumed that the patching functions $F_\mu$ are
functions of only two variables in the corresponding plane
${\bf r}_\mu={\bf L}_\mu$, $\mu=x$, $y$, or $z$. 

\begin{figure}[htbp]
  \begin{center}
    \leavevmode
    \epsfbox{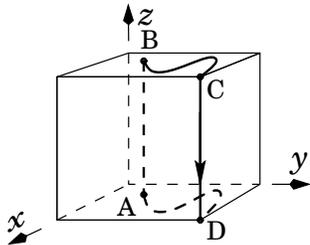}
    \caption{The coupling of the gauge field with the current line
      $\rm AB$ is made gauge-invariant by introducing the surface
      patching term (\protect\ref{integral-through}), equivalent to
      integration along the contour $\rm BCDA$.  The line $\rm BC$ is
      the image of $\rm AD$ after translation by the period
      ${\bf{}L}_z$. 
      }
    \label{fig:cub-path}
  \end{center}
\end{figure}

Generally, under the gauge transformation~(\ref{gauge-transf}) the
fields ${\bf A}$ at the opposite faces of the system transform
differently, so that the values of the increments $\Incr_\mu{\bf
  A}_\perp$ change.  To preserve the
condition~(\ref{patching-f-in-twod}), we need to adjust the patching
functions, $ F_\mu\rightarrow F_\mu+\Incr_\mu f $.  This
transformation rule, along with the original
definition~(\ref{patching-f-in-twod}), can be satisfied by defining
\begin{equation}
  \label{integral-through}
  F_\mu({\bf r}_\perp)= %
  \int\limits_{\rm BCDA} %
  {\bf A}({\bf r})\cdot d{\bf r}+{\rm Const},
\end{equation}
where the integration is performed along the three-part contour
starting in the point ${\rm B}\equiv{\bf r}_\perp+{\bf L}$ and ending
in ${\rm A}\equiv{\bf r}_{\perp}$ as shown in FIG.~\ref{fig:cub-path}:
two congruent lines BC and DA along the opposite faces of the system,
connected by a {\em fixed\/} contour CD entering both faces in
equivalent points.  An arbitrary gauge-invariant constant introduced
in Eq.~(\ref{integral-through}) accounts for the freedom of uniformly
``twisting'' the boundary conditions for $\theta$; it is also the
freedom of choosing the configuration of the fixed segment CD of the
``back flow'' contour BCDA.

\subsection{2D example}
Before advancing any further the analysis of the three-dimensional
problem, let us analyze a simpler two-dimensional Villain
\xy\ model
\begin{equation}
  \label{Villain-2d-xy}
  i {\cal S}_{\rm V}= %
  {1\over2T}\sum_{\bf r}{
    \left(\nabla\theta-{\bf A}-2\pi{\bf m}\right)^2}
\end{equation}
in external magnetic field.  In
expression~(\ref{Villain-2d-xy}) the angles $\theta$ are defined in
the vertices of two-dimensional square lattice, the vector
potential of the external magnetic field is defined at the links, as
well as the additional integer-valued field ${\bf m}$ which restores
the periodicity broken after changing the cosines of the
original \xy\ model for parabolas in the Villain approximation.

At every link the summation in appropriate component of ${\bf m}$ is
independent of anything else and can be transformed with the Poisson
summation formula, 
\begin{equation}
  \label{vxy-poisson-transformed}
  {1\over2T}  \left(\nabla\theta-{\bf A}-2\pi{\bf m}\right)^2
  \longrightarrow
  {T{\bf j}^2\over2}-i{\bf j}\cdot \left(\nabla \theta-{\bf A}\right).
\end{equation}
replacing the summation over ${\bf m}$ by the summation over new
independent integer-valued variables ${\bf j}$ also defined
at every link of the lattice.

\begin{figure}[htbp]
  \begin{center}
    \leavevmode
    \epsfxsize=1.5in
    \epsfbox{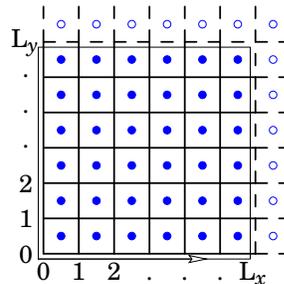}
    \caption{Solid lines denote the definition region of the 2D Villain
      \xy\ model, while the filled circles show the corresponding
      vertices of the dual lattice.  Region outside the thin solid
      contour represent the images of the points after
      translation(s).}
    \label{fig:sq-lat}
  \end{center}
\end{figure}

Normally, the term with the gradient of $\theta$ in
Eq.~(\ref{vxy-poisson-transformed}) can be integrated by parts,
$-i{\bf j}\cdot \nabla \theta\rightarrow i\theta \nabla\cdot{\bf j}$, %
so that the subsequent integration in $\theta$ renders the
conservation law
\begin{equation}\label{conservat-of-j}
  \nabla\cdot{\bf j}=0,
\end{equation}
which, given the form of its coupling to the gauge field, allows to
identify ${\bf j}$ as the conserved current.  In the finite system,
however, the phases $\theta$ in equivalent points across the boundary
of the system are related by the patching
equations~(\ref{patching-cond}), and the integration by parts
generates additional boundary terms
\begin{equation}
  i{\cal S}_{\rm surf} %
  = -i\int_0^{{\bf L}_y}  {\bf j}^x\cdot F_x(y) dy %
    -i\int_0^{{\bf L}_y}  {\bf j}^y\cdot F_y(x) dx, %
  \label{surface-term-2d}
\end{equation}
where the patching functions $F_i$ are related to the increments of
the gauge field across the system,
\begin{equation}\label{increments-2d}
  \Incr_x {\bf A}_y=\nabla_y F_x(y),\qquad
  \Incr_y {\bf A}_x=\nabla_x F_y(x).
\end{equation}
As before, these equations define the functions $F_x$ and $F_y$ up to
arbitrary additive constants.

The term with the square of the current ${\bf j}$ in
Eq.~(\ref{vxy-poisson-transformed}) can be split by the
Hubbard-Stratonovich transformation,
\begin{displaymath}
  {T {\bf j}^2\over2}\rightarrow i {\bf b} {\bf j}+{{\bf b}^2\over2T},
\end{displaymath}
where ${\bf b}$ is the new auxiliary field defined at the links of the
lattice.  Although this field couples with the current like an
additional gauge field, the integration in ${\bf b}$ is performed over
all possible values including different gauges; this field is not a
gauge field.  By construction, the field ${\bf b}$ is defined only
within the period of the system and, therefore, it satisfies periodic
boundary conditions and does not introduce any non-trivial total
magnetic flux.

Now let us resolve the conservation law~(\ref{conservat-of-j}) by
introducing an integer-valued scalar field $h$ associated with the
vertices of the dual lattice
$  %
{\bf j}^\mu=\epsilon^{\mu\nu}\nabla_\nu h\equiv \widetilde\nabla^\mu h
$. %
This definition implies that the increments of the additional field
$h$ across the system are related to the total currents in
perpendicular directions, 
\begin{equation}\label{eq:2d-discont}
  \Incr_x h=-I_y \quad {\rm and}\quad \Incr_y h=I_x;
\end{equation}
because of the periodicity and the current conservation these
increments remain {\em constant\/} along the edges.

At this point  the effective action can be written as
\begin{equation}
  \label{vxy-2d-one}
  i{\cal S}_{\rm V}= \int \left\{ %
    {{\bf b}^2\over2T} 
    +i ({\bf b}+{\bf A})\widetilde\nabla h 
  \right\} dS 
  +i{\cal S}_{\rm surf},
\end{equation}
where the surface patching term ${\cal S}_{\rm surf}$ is given by
Eq.~(\ref{surface-term-2d}).  To arrive at the dual representation of
the model we need to integrate the gauge coupling by parts, using the
identity\footnote{The product of the lattice fields ${\bf A} h$ is
  defined by taking the field $h$ at the plaquette shifted in the
  positive direction (up or right) of the link with the field ${\bf
    A}$}
\begin{displaymath}
  {\bf A}_\mu\epsilon^{\mu\nu}\nabla_\nu h
  \equiv{\bf A}\!\times\!\nabla h
%  =h\nabla_\mu\epsilon^{\mu\nu}{\bf A}-\nabla_\mu\epsilon^{\mu\nu}({\bf A}h)
  =h\,\nabla\!\times\!{\bf A}-\nabla\!\times({\bf A}h).
\end{displaymath}
The last term of this expression results in a contour integral along
the edge of the system, which can be split into vertical and
horizontal parts,
\begin{equation}
  \label{contour-integral}
  -\oint  h \,{\bf A}\, d{\bf l}= %
     \int_0^{{\bf L}_x}\!\!\Incr_y ({\bf A}_x h ) dx %
  -\!\int_0^{{\bf L}_y}\!\! \Incr_x ({\bf A}_y h ) dy.
\end{equation}
With the increments of fields ${\bf A}$ and $h$ known, we can
transform the increments of their products using the expression
\begin{equation}\label{difference}
  \Incr_x (f g) % = f Dg +g Df -Df Dg
  =f({\bf L}_x,y) \Incr_x g %
  +g({\bf L}_x,y)\Incr_x f %
  -\Incr_x f \Incr_x g,
\end{equation}
so that the contour integral~(\ref{contour-integral}) takes the form
\begin{eqnarray}
  \label{contour-integral-two}
  -\oint h\,{\bf A} \, d{\bf l}&=& %
  \int_0^{{\bf L}_x}\! \left( %
    \nabla_x F_y \, h +I_x {\bf A}_x-I_x \nabla_x F_y %
  \right) dx\\ %
  &-& \int_0^{{\bf L}_y}\! \left( %
    \nabla_y F_x \, h -I_y {\bf A}_y+I_y \nabla_y F_x %
  \right) dy.  \nonumber
\end{eqnarray}
The first term in each of these integrals can be integrated by parts,
with the integrated part once again evaluated using
Eq.~(\ref{difference}).  With the expression for the total
flux of the magnetic field, $\Phi=\Incr_y
F_x-\Incr_x F_y$, following immediately from the Stokes formula and
the definition~(\ref{increments-2d}), we obtain
\begin{eqnarray}
  \label{contour-integral-fin}
  -\oint h\,{\bf A}\,d{\bf l}&=& %
  \int_0^{{\bf L}_y}\!{\bf j}_x F_x dy +\int_0^{{\bf L}_x}\!{\bf j}_y F_y dx %
  - h({\bf L}_x,{\bf L}_y)\Phi %
  \nonumber \\ %
  &\relax&+I_x\left\{\int_0^{{\bf L}_x}\!\! {\bf A}_x dx+\Phi-F_x({\bf L}_y)\right\} %
  \nonumber \\ %
  &\relax&+I_y\left\{\int_0^{{\bf L}_y}\!\! {\bf A}_y dx-\Phi-F_y({\bf L}_x)\right\}.%
\end{eqnarray}
The first two terms cancel exactly with the edge patching
term~(\ref{surface-term-2d}), while the remaining terms can be
eliminated by properly defining the so far arbitrary additive
constants for the scalar field $h$ and the patching functions $F_i$.
Specifically, defining $h({\bf L}_x,{\bf L}_y)=0$,
\begin{displaymath}
  F_x({\bf L}_y)=\int_0^{{\bf L}_x}A_x(x,{\bf L}_y)dx+\Phi
\end{displaymath}
and 
\begin{displaymath}
  F_y({\bf L}_x)=\int_0^{{\bf L}_y}A_y({\bf L}_x,y)dy-\Phi,
\end{displaymath}
eliminates the last two terms in Eq.~(\ref{contour-integral-fin}), so
that the performed integration by parts can be written simply as
\begin{displaymath}
  \int A\times\!\nabla h \,dS\,+\,{\cal S}_{\rm surf}= %
  \int h\,\nabla\times  {\bf A}\, d{\bf s}. %
\end{displaymath}
%Here ${\bf a}={\bf A}+{\bf b}$; because of the periodicity of the
%Hubbard-Stratonovich field ${\bf b}$ we do not have to modify the
%patching functions $F_i$.  
Although the described procedure works generally for any gauge field
${\bf A}$, we had to be careful to separate the contribution from the
Hubbard-Stratonovich field ${\bf b}$: the gauge-fixing terms cannot
depend on this field.  The field ${\bf b}$ is periodic and the
equivalent series of transformations is much simpler, immediately
leading to 
\begin{equation}
  \label{eq:2d-b-contrib}
  -\oint h\,{\bf b}\,d{\bf l}
  =I_x\!\int_0^{{\bf L}_x}\! {\bf b}_x dx
  +I_y\!\int_0^{{\bf L}_y}\! {\bf b}_y dy
\end{equation}
instead of Eq.~(\ref{contour-integral-fin}).  It is convenient to use
the potential representation 
$ %
  {\bf b}_i =\nabla_i f+ \epsilon^{ij}\nabla_j g
$ %
with the condition that the normal
derivatives of both potentials vanish at the boundary,
$ %
  \partial g/\partial n=\partial f/\partial n=0
$.  %
Then it follows from the periodicity
\begin{displaymath}
  \int {\bf b}^2\, dS=\int (\nabla f)^2+(\nabla g)^2 \, dx\,dy, 
\end{displaymath}
and the transformed action~(\ref{vxy-2d-one}) can be written as
\begin{eqnarray*}
  i{\cal S}_{\rm V}=\!\int\!\! \left\{{(\nabla g)^2\over 2T} %
    + ih\,(\nabla\!\times\!{\bf A}-\nabla^2 g) \right\}dS %
  + \!\int\!\! {(\nabla f)^2\over2T}d S\\ %
  + i(I_x+I_y)\,f({\bf L}_x,{\bf L}_y)  -i I_x \, f(0,{\bf L}_y)  -i I_y \, f({\bf L}_x,0),
\end{eqnarray*}
with the free terms in the last line resulting from the line
integrals~(\ref{eq:2d-b-contrib}).  It is clear that the two
potentials $f$ and $g$ in this action are completely independent of
each other.  The summation over integer $h$ only fixes the vortex
charges $\upsilon=\nabla\!\times\!{\bf A}-\nabla^2 g$ to integer
values modulo $2\pi$.  Since the field ${\bf b}$ is single-valued, it
does not contribute any additional flux and the total vorticity must
be exactly equal to the total flux of the external magnetic field.

Furthermore, the term $\int (\nabla g)^2 dS /2T$  precisely equals to
the Coulomb energy associated with the vortex charges in the uniform
background.  This can be proven by introducing an auxiliary field
$\phi$,
\begin{eqnarray*} 
  {(\nabla g)^2\over 2T}\rightarrow  %
  {T (\nabla \phi)^2\over2}+i\nabla  \phi \nabla g \rightarrow %
  {T (\nabla \phi)^2\over2}-i\phi\nabla^2 g\\  %
  ={T(\nabla\phi)^2\over2}-2\pi i\phi\upsilon 
  +i\phi{\nabla\!\times\!{\bf A}},
\end{eqnarray*} 
where the integrated part vanished because of the boundary condition
$\partial g/\partial n=0$.  

The considered transverse part of the dual action is independent of
the lateral currents $I_x$, $I_y$ encoded in the discontinuity of the
boundary condition~(\ref{eq:2d-discont}) for the field $h$.  These
currents act as the only sources of the scalar potential $f$.  The
energy of the equivalent Coulomb problem, quadratic in these currents,
provides the only current-dependent contribution to the free energy;
this contribution is obviously non-critical.

%%%Another interesting case is the lattice superconductor, where the
%%%original gauge field ${\bf A}$ has some dynamics, for example, the
%%%usual Maxwell term $(\nabla\times {\bf A})^2/2e^2$.  Then, after using
%%%the ``scalar'' potential $f$ to separate the lateral currents, the
%%%remaining gauge field 
%%%\begin{math}
%%%  {\bf a}={\bf A}+\widetilde\nabla g %={\bf A}+{\bf b}'
%%%\end{math}
%%%has the effective dynamics described by the coupling
%%%\begin{displaymath}
%%%  {\bf a}\,
%%%  {P^2\over P^2 + {e^2/T}}\,
%%%  {\bf a}, 
%%%\end{displaymath}
%%%so that the model once again can be formulated in terms of the scalar
%%%integer-valued field $\upsilon=\nabla\!\times\!{\bf a}$, with the
%%%interaction $v_1G(r_{12})v_2$, where
%%%\begin{displaymath}
%%%  G(r)=\int {d^2k\over (2\pi)^2 }
%%%  {e^{i\bf kr}\over {e^2/T}+{\bf P}^2}
%%%\end{displaymath}
%%%describes the screened at large distances interaction between the
%%%vortices, and ${\bf P}_i=2\sin(k_i/2)$ is the usual lattice momentum.
%%%As in the usual \xy\ model, the average magnetic field %
%%%$\langle{\bf B}\rangle$ determines the number density of the vortices
%%%in the system.

\section{Duality in three dimensions}
\label{sec:3d-duality}

Now let us resume the analysis of the Chern-Simons gauge model in
three dimensions.  The major steps include:
(\ref{par:cs-3d-defnd})~Start with the canonical
form~(\ref{eq:def-csxy}) of the model and rewrite it in terms of
conserved integer-valued current ${\bf j}$ and auxiliary field ${\bf
  b}$.  (\ref{par:curr-3d-conserv})~Resolve the current conservation
law by introducing an additional integer-valued field ${\bf m}$, ${\bf
  j}=\nabla\!\times\!{\bf m}$, integrate the current coupling by parts
and prove that, as a result of summation in ${\bf m}$, the flux of the
total gauge field through every plaquette is an integer, identified
with the dual conserved current.  Up to this point all transformations
will be done {\em exactly\/} for any form of the original \xy\ model.
(\ref{par:3d-scalar-poten})~In the Villain approximation, integrate
over the longitudinal part of the field ${\bf b}$ to allow treating of
its remaining part as a gauge field.
(\ref{par:constr-3d-resolv})~Finally, resolve the constraint on the
original Chern-Simons gauge field by yet another Hubbard-Stratonovich
transformation, and express the result in the canonical
form~(\ref{eq:def-csxy}).

\paragraph{Introduce the integer-valued current:}
\label{par:cs-3d-defnd}
Before introducing the surface patching terms for the
model~(\ref{eq:def-csxy}), it is convenient to transform its scalar
part~(\ref{eq:def-scalar}) to the equivalent current representation
\begin{equation}
  \label{eq:3d-cs-defnd-curr}
  i{\cal S}_\theta= %
  i\sum{\bf j}\cdot %
  \left(\nabla \theta\!-\!{\bf a}\!-\!{\bf A}\!-\!{\bf b}\right) %
  -\sum{\cos{\bf b}\over T},
\end{equation}
where the original form~(\ref{eq:def-scalar}) is restored after the
sum\-ma\-tion\footnote{The summation in discrete values of ${\bf j}$
  automatically accounts for the periodicity of the cosine, validating
  its expansion in powers of ${\bf b}$.  For example, the quadratic
  approximation $\cos{\bf b}\rightarrow 1-{\bf b^2}/2$ is equivalent
  to the na\"\i{}ve Villain approximation for the original
  model~(\protect\ref{eq:def-scalar}).  The regular expansion in
  powers of~${\bf b}$ can be constructed by analytically continuing
  the Fourier coefficients $I_m(T^{-1})$ of $\exp[(\cos b)/T]$ to all
  real values of $m$, and using the function
  $ %
  \chi (b)=T\ln\int_0^{\infty} I_\lambda(T^{-1}) \cos(\lambda b)
  d\lambda/(2\pi)
  $ %
  instead of $\cos b$ in the exponent of
  Eq.~(\protect\ref{eq:3d-cs-defnd-curr}).} %
over all integer ${\bf j}$ and subsequent integration over the
auxiliary field ${\bf b}$ at every link.  As in the two-dimensional
example, the integration over the phases $\theta$ imposes the
conservation constraint $\nabla\cdot{\bf j}=0$, so that ${\bf j}$ can
be identified as the integer-valued current.

\begin{figure}[htbp]
  \begin{center}
    \leavevmode
    \epsfbox{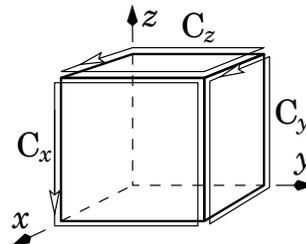}
    \caption{The integration areas ${\rm S}_i$ in the surface patching
      terms~(\protect\ref{eq:3d-cs-surface-terms}) are limited by the
      contours ${\rm C}_i$, $i=x$, $y$, $z$.}
    \label{fig:cube}
  \end{center}
\end{figure}

Assuming the kernel ${\cal K}$ is transverse ({\em i.e.}\ $({\cal
  K}{\bf a})$ gauge-invariant,) it is easy to find the required
  form 
\begin{equation}
  \label{eq:3d-cs-surface-terms}
  i{\cal S}_{\rm surf}=i\sum_i\int_{S_i} %
   F_i\,\left({\bf j}-{1\over2}{\cal K}{\bf a}\right) \cdot d{\bf s}
\end{equation}
of the surface terms required to fix the gauge-invariance in the
finite system.  Here $F_i$ are the patching functions for the sum of
two gauge fields ${\bf a}_{\rm tot}={\bf a}+{\bf A}$ as defined in
Appendix~\ref{appen:hermiticity}, and the integration areas ${\rm
  S}_i$ are the three visible faces of the cube as shown in
FIG.~\ref{fig:cube}.  Since the auxiliary field ${\bf b}$ is
single-valued, it creates no total magnetic flux through the system
and there are no boundary patching terms associated with this field.

The surface patching terms~(\ref{eq:3d-cs-surface-terms})
make the model gauge-invariant, but they
also eliminate the integration over the constant part of the
fluctuating gauge field ${\bf a}$ along with the resulting mean-field
constraint 
\begin{equation}
  \label{eq:3d-mf-constraint}
  \int {\cal K}{\bf a}\,d\Omega=-i\int {\bf j}\,d\Omega.
\end{equation}
For the model with purely Chern-Simons gauge
kernel~(\protect\ref{eq:def-purely-cs}),
Eq.~(\protect\ref{eq:3d-mf-constraint}) is the condition on the
average flux of the associated magnetic field
$\langle\nabla\!\times\!{\bf a}\rangle %
=2\pi\alpha\,\langle{\bf j}\rangle$.  This constraint allows us to
associate the model~(\ref{eq:def-csxy}) with the continuum CSLG
model~(\ref{eq:def-cslg}) and ultimately with the real quantum Hall
systems; therefore the constraint~(\ref{eq:3d-mf-constraint}) needs to
be imposed on the original model~(\ref{eq:def-csxy}) by integration
over an additional uniform auxiliary field.
%%% This is equivalent to the
%%%transformation between the big and small canonical ansamble by fixing
%%%the number of particles with a fluctuating chemical potential.

\paragraph{Resolve the current conservation:}
\label{par:curr-3d-conserv}
By analogy with the two-dimensional problem, we can resolve the
lattice current-conservation constraint $\nabla\!\cdot{\bf j}=0$ by
introducing an auxiliary field ${\bf m}$, 
\begin{equation}
  \label{eq:3d-current-split}
  {\bf j}=\nabla\!\times{\bf m}.
\end{equation}
There is, however, an important difference with the two-dimensional
case: in 3D the gauge freedom associated with the solution of
Eq.~(\ref{eq:3d-current-split}) is much more broad.  Simultaneously,
in generic gauge ${\bf m}$ is {\em not\/} an integer-valued field, and
even the existence of the integer-valued solution is not immediately
obvious.

Let us choose the gauge $m_z=0$, then it is easy to check that the sums
\begin{eqnarray}
  \label{eq:3d-current-solution}
  m_x(x,y,z)&=& \int_0^z\!\! j_y(x,y,\zeta) d\zeta %
  -\int_0^x j_z(\xi,y,0) d\xi, %
  \nonumber \\
  m_y(x,y,z)&=& -\int_0^z \!\! j_x(x,y,\zeta) d\zeta
\end{eqnarray}
satisfy Eq.~(\ref{eq:3d-current-split}); by construction this solution
is integer-valued.  Now we may introduce the patching functions $G_i$
associated with the integer-valued gauge field ${\bf m}$.  Explicit
calculation shows that both $G_x=0$ and $G_y=0$ up to an integer
constant, and
\begin{displaymath}
  \partial_x G_z(x,y)=\int_0^{{\bf L}_z} j_y dz,\qquad
  \partial_y G_z(x,y)=-\int_0^{{\bf L}_z} j_y dz;
\end{displaymath}
these equations have a solution since the curl 
\begin{eqnarray*}
  \partial_y(\partial_x G_z)-\partial_x(\partial_y G_z)
  &=&\int_0^{{\bf L}_z}\!\left(\partial_x j_x+\partial_y j_y\right)\,dz\\
  &=&-\int_0^{{\bf L}_z}\! \partial_z j_z \,dz=0
\end{eqnarray*}
vanishes.  According to Eq.~(\ref{eq:cs-hermiticity}), the complete
current-dependent coupling can be integrated by parts as
\begin{eqnarray}
  \label{eq:3d-current-by-parts}
  \lefteqn{\int\!\nabla\!\times\!{\bf m}\cdot %
    ({\bf b}+{\bf a}_{\rm tot})\,d\Omega %
    +{\cal S}_{\rm surf} %
    =\int\! {\bf m}\cdot\nabla\!\times\!({\bf b}+{\bf a}_{\rm
      tot})\,d\Omega
    }&\relax&\nonumber\\
  &\relax&+\sum_{i=x,y,z}I_i\int_0^{{\bf L}_i}\! {\bf b}_i\, dx_i
  + \int_{S_z}\!\! G_z \left[\nabla\times\!({\bf b}+{\bf a}_{\rm
      tot})\right]_z\,d{\bf s},
  % \hskip3in 
\end{eqnarray}
where ${\bf a}_{\rm tot}={\bf a}+{\bf A}$.   Independent summations
over $m_x$ and $m_y$ restrict the $x$- and $y$- components of the
field
\begin{equation}
  \label{eq:3d-dual-curr}
  {\bf v}= %
  {1\over2\pi}\nabla\times\!({\bf b}+{\bf a}_{\rm tot})
\end{equation}
to integer values.  This field is a total curl, and consequently it is
conserved, $\nabla\!\cdot\!{\bf v}=0$.  Therefore, the differences
$\partial_z{\bf v}_z$ of the $z$-component of this field at adjoining
links are also integer-valued.  Finally, the summation of the patching
surface term at the surface $S_z$ over the values of the function
$G_z$ restricts the values of ${\bf v}_z$ at this surface to integer
values, completing the proof that all components of the field ${\bf
  v}$ can take only integer values.  This conserved field ${\bf v}$
will be identified as the dual integer-valued current.

Up to this point all transformations were exact and very general,
valid for virtually any imaginable form of the lattice gauge model.
Eq.~(\ref{eq:3d-dual-curr}) relates the lines of the dual current to
the flux lines of the combined magnetic field $\nabla\!\times\!({\bf
  A}+{\bf a}+{\bf b})$, so that the total dual current through the
system
\begin{equation}
  \label{eq:total-dual-current}
  \widetilde{\bf I}=\int {\bf v}\,d{\bf s} %
  ={1\over2\pi}\int \nabla\!\times\!({\bf A}+{\bf a}) \,d{\bf s} %
  ={{\bf \Phi}\over2\pi}
\end{equation}
in any given direction is precisely equal to the total number of flux
quanta ${\bf \Phi}/2\pi$ of the combined magnetic field in this
direction (the contribution of the periodic auxiliary field ${\bf b}$
vanishes after the surface integration).  This universal relationship
between the total current and the total magnetic flux in mutually dual
models is one of the central results of this work.  Previously, a
similar result, relating the flux of the magnetic field with the
number of vortices, was obtained by Lee and
Fisher\cite{Fisher-89,Lee-89}, in an approximate lattice duality
involving a ``softening'' of the integer constraint on the total
number of vortices.

\paragraph{Integrate away the scalar potential:}
\label{par:3d-scalar-poten}
The goal of the subsequent steps is to rewrite the dual action in a
form possibly close to that of the original
action~(\ref{eq:def-scalar}).  This can be done only approximately,
ignoring some terms irrelevant in the RG sense.  In particular, the
integration over the auxiliary gauge field ${\bf b}$ will be done only
in the Villain approximation, leaving out the non-linear corrections.
With this in mind, the result of the summation of the combined
action~(\ref{eq:def-gauge-part}),~(\ref{eq:3d-cs-defnd-curr}) over the
integer-valued current ${\bf j}$ can be written as
\begin{eqnarray}
  \label{eq:3d-dual-fixed-lagr}
  i{\cal S}_{\rm cs}&\rightarrow& \int {{\bf
      b}^2\over2T}d\Omega+{1\over2}\int{\bf a}{\cal K}{\bf a}\,d\Omega
  \nonumber\\  &\relax&   
  +i\sum_{\mu=x,y,z}I_\mu\int_0^{{\bf L}_\mu}\!{\bf b}_\mu\, dx_\mu %
  + (\text{surface terms}).
\end{eqnarray}
Here the field ${\bf a}$ is a solution of the equation
\begin{equation}
  \label{eq:3d-constraint}
  \nabla\!\times\! {\bf a} %
  =2\pi{\bf v}-\nabla\!\times\!({\bf b}+{\bf A}),
\end{equation}
while the Hubbard-Stratonovich field ${\bf b}$ and the conserved dual
integer-valued current ${\bf v}$ with given total flux $\int {\bf
  v}\,d{\bf s}={\bf \Phi}/2\pi$ through the sample are independent
variables.  The presence of surface terms ensure that the
expression~(\ref{eq:3d-dual-fixed-lagr}) does not depend on the
particular gauge chosen for the solution ${\bf a}$ of
Eq.~(\ref{eq:3d-constraint}).

Following the two-dimensional example, we can eliminate the integrals
of ${\bf b}$ in the second line of Eq.~(\ref{eq:3d-dual-fixed-lagr})
by writing ${\bf b}=\nabla \phi+\nabla\!\times\! {\bf
  g}=\nabla\phi+{\bf b}'$, with the tangential part of the second term
vanishing everywhere at the surface of the system.  Then the
integration over the field $\phi$ yields a non-critical quadratic in
the currents ${\bf I}_i$ contribution to the free energy, proportional
to the energy of the Coulomb interaction of the charges $\pm {\bf
  I}_i$ set in the corners of the cube in FIG.~\ref{fig:cube} and
their mirror images.

Now the constraint on the boundary values of the field ${\bf b}'$ can
be removed by adding the appropriate surface patching terms rendering
the expression gauge-independent; this is possible because the surface
terms evaluate to zero with the original
boundary condition of vanishing %${\bf b}'_{\rm t}=0$ for the
tangential components of the field ${\bf b}'$.  The gauge fields ${\bf
  a}$ and ${\bf b}'$ enter Eq.~(\ref{eq:3d-constraint}) only in the
combination ${\bf a}'= {\bf a}+{\bf b}'$; integrating away their
difference we obtain the combined gauge coupling\cite{Pryadko-95}
\begin{displaymath}
  {({\bf b}')^2\over2T}+  {1\over2}{\bf a}{\cal K}{\bf a}
  \rightarrow  {1\over2}{\bf a}'{\cal K}'{\bf a}', 
  \qquad {\cal K}'=\left(T+{\cal K}^{-1}\right)^{-1},
\end{displaymath}
where the field ${\bf a}'$ satisfies the equation
\begin{displaymath}
  \nabla\!\times\!{\bf a}'=2\pi{\bf v}-\nabla\!\times\!{\bf A},
\end{displaymath}
and all inverse matrices must be evaluated in the transverse gauge
$\nabla\cdot{\bf a}=0$. 

\paragraph{Introduce new auxiliary field:}
\label{par:constr-3d-resolv}

Finally, we need to express the coupling ${\bf a}'{\cal K}'{\bf a}'$
as a path integral over an additional auxiliary field ${\bf c}$.  With
the appropriate surface patching integrals implicit, this can be done
by the Hubbard-Stratonovich transformation
\begin{equation}
  \label{eq:hubbard}
  {1\over2}{\bf a}'{\cal K}'{\bf a}'\rightarrow 
  i{\bf a}'\cdot\nabla\!\times\!{\bf c} %
  + {1\over2}(\nabla\!\times\!{\bf c})
  {{\cal K}'}^{-1}(\nabla\!\times\!{\bf c}),
\end{equation}
where the additional constraint on the total flux of the auxiliary
field through the system
\begin{equation}
  \label{eq:3d-aux-flux}
  \int\nabla\!\times\!{\bf c}\,d{\bf s} %
  =-i\int {\cal K}' {\bf a}' \,d{\bf s}=-{\bf I}
\end{equation}
is mandated by the degeneracy of the gauge kernel ${\cal K}'$ at zero
momentum due to the presence of the surface patching terms.  Indeed,
the Hubbard-Stratonovich transformation~(\ref{eq:hubbard}) can be
interpreted as the addition in the exponent 
\begin{eqnarray*}
  {\bf a}' {\cal K}' {\bf a}'\rightarrow {\bf a}' {\cal K}' {\bf a}'+
  \left(\nabla\!\times\!{\bf c}+i{\cal K}'{\bf a}'\right) {{\cal K}'}^{-1}
  \left(\nabla\!\times\!{\bf c}+i{\cal K}'{\bf a}'\right),
\end{eqnarray*}
with the subsequent integration in~${\bf a}'$; the
condition~(\ref{eq:3d-aux-flux}) is required for the inverse operator
to be meaningful.

With the help of Eq.~(\ref{eq:3d-constraint}), the total Lagrangian
becomes
\begin{equation}
  \label{eq:3d-dual-lagr-one}
  i\widetilde{\cal L}_{\rm cs} %
  =i{\bf c}\cdot(2\pi{\bf v} - \nabla\!\times\!{\bf A}) %
  + {(\nabla\!\times\!{\bf c}) %
  {{\cal K}'}^{-1}(\nabla\!\times\!{\bf c})\over2}. %
\end{equation}
After defining the dual gauge field
\begin{displaymath}
  \nabla\!\times\!\tilde{\bf a} %
  = 2\pi \nabla\!\times\!{\bf c}-2\pi i\,{\cal K}'{\bf A}, %
\end{displaymath}
we can write the Lagrangian of the system in the form
\begin{equation}
  \label{eq:3d-dual-lagr}
  i\widetilde{\cal L}_{cs}  = {{\bf A}{\cal K}'{\bf A}\over2} %
  + i{\bf v} \left( %
    \tilde{\bf a} -2\pi  \,(i\nabla\times)^{-1}{\cal K}'\,{\bf A} %
  \right)
  + {\tilde{\bf a}\,\widetilde{\cal K}'\,\tilde{\bf a}\over2}, 
\end{equation}
where the dual gauge kernel 
\begin{displaymath}
  \widetilde{\cal K} %
  = {{\bf P}\times{{\cal K}'}^{-1}\times{\bf P}\over(2\pi)^2} %
  = {T P^2\over(2\pi)^2}
  -{{\bf P}\times{\cal K}^{-1}\times{\bf P}\over(2\pi)^2},
\end{displaymath}
and the total flux $\widetilde\Phi$ of the gauge fields coupled to the
dual current ${\bf v}$ is determined by the original particle current,
\begin{equation}
  \label{eq:3d-dual-flux}
  \widetilde\Phi = \int\left(
    \nabla\!\times\!  \tilde{\bf a}+2\pi i{\cal K}'A %
  \right)\,d{\bf s} %
  = -2\pi{\bf I}
\end{equation}
as immediately follows from Eq.~(\ref{eq:3d-aux-flux}) and the
definition of the dual gauge field $\tilde{\bf a}$.  It is easy to see
that the second duality transformation restores the form of the model
up to the trivial symmetry~(\ref{eq:inversion}).  

In the limit of small momenta or slow-varying external field ${\bf
  A}$, and assuming the original gauge kernel~(\ref{eq:def-genrl-cs})
has purely Chern-Simons form~(\ref{eq:def-purely-cs}), the important
part of the dual model can be written as
\begin{equation}
  \label{eq:3d-dual-lagr-simp}
  \widetilde{\cal L}_{\rm cs} \sim 
   {\bf v} \left( %
    \tilde{\bf a} +{1\over\alpha}\,{\bf A} %
  \right)
  +{\alpha\over4\pi} ({\tilde{\bf a}\partial\tilde{\bf a}})
  +i{T\over4\pi^2}(\nabla\!\times\!\tilde{\bf a})^2, 
\end{equation}
indicating that, up to irrelevant difference in the quasiparticle's
size, the dual model is the Chern-Simons Villain \xy\ model with the
excitations of electric charge $1/\alpha$, statistics $-1/\alpha$,
and the dimensionful lattice temperature\cite{Pryadko-95,Fradkin-96}
$\tilde{T}=T/\alpha^2$.   

The global conditions on the total dual flux~(\ref{eq:3d-dual-flux})
and current~(\ref{eq:total-dual-current}) restore the mean-field
constraint~(\ref{eq:3d-mf-constraint}) for the dual model only if the
gauge kernel of the original model has purely Chern-Simons
form~(\ref{eq:def-purely-cs}).  Indeed,  
\begin{eqnarray*}
 \int\!\nabla\!\times\!\tilde{\bf a}\,d{\bf s} %
  &\stackrel{(\ref{eq:3d-dual-flux})}{=}&
  \widetilde{\bf\Phi}-{{\bf\Phi}_A\over\alpha} %
  =-2\pi{\bf I} -{{\bf\Phi}_A\over\alpha}\\
  &=&
  -{2\pi\over\alpha}\left(\alpha{\bf I}+{{\bf\Phi}_A\over2\pi}\right)
  =-{1\over\alpha}{\bf\Phi}
  \stackrel{(\ref{eq:total-dual-current})}{=}
  -{2\pi\over\alpha}\tilde {\bf I},
\end{eqnarray*}
where ${\bf \Phi}_A$ denotes the total flux of the external magnetic
field.  It is this property that shows that the chosen definition of
the gauge coupling in finite systems is natural and corresponds to the
expected behavior for the model describing quantum Hall systems.  The
relationship of the considered class of models to the quantum Hall
effect will be discussed more specifically in
Sec.~\ref{sec:correspondence}.

Unfortunately, it is not clear how to perform the transition between
Eqns.~(\ref{eq:3d-dual-lagr-one}) and~(\ref{eq:3d-dual-lagr}) for the
generalized form of the Chern-Simons
coupling~(\ref{eq:gen-cs-kernel}), which is supposed\cite{Fradkin-96}
to describe the even-denominator (metallic) fixed points of the
quantum Hall system.  In this case the background charge density
fluctuations in Eq.~(\ref{eq:3d-dual-lagr-one}) cannot be
self-consistently traded for the modified magnetic field in
Eq.~(\ref{eq:3d-dual-lagr}), indicating that additional entities like
the ``magnetic'' charges\cite{Fradkin-96} are needed.  At this time,
it is not clear what physical meaning can be attributed to such
objects.

\section{Flux attachment transformation}
\label{sec:fa-transform}

In addition to the duality, the continuum Chern-Simons models have
another important symmetry, the periodicity, or the symmetry under the
flux attachment transformation.  In this section I show that the
lattice version of this transformation is exact and, in addition to
modifying the Chern-Simons coefficient\cite{Pryadko-95}, changes the
magnetic flux ${\bf \Phi}$ by the amount proportional to the global
current ${\bf I}$.

Let us fix a configuration of integer-valued conserved current %
${\bf j}$ and define the related gauge field ${\bf c}_{\bf j}$ as a
solution of the equation $\nabla\!\times\!{\bf c}_{\bf j}=4\pi{\bf
  j}$.  In a system where the current lines never cross the boundary,
the coupling
\begin{equation}
  \label{eq:fa-self-coupling}
  X={1\over2}\int {\bf c}_{\bf j}\,{\bf j}\, d\Omega %
  +\text{surface terms}
\end{equation}
is always an integer times $2\pi$.  Because of the surface patching
terms, this coupling is always an integer times $2\pi$, independent of
the gauge chosen for the field ${\bf c}_{\bf j}$.  Indeed, for the
current lines that do cross the boundary, the surface terms can be
interpreted as the ``back flow'' currents, as shown in
FIG.~\ref{fig:cub-path}, and the statement for the closed current
lines apply.

We would like to rewrite the coupling~(\ref{eq:fa-self-coupling}) as a
path integral over a dynamical variable ${\bf c}$ instead,
\begin{equation}
  \label{eq:fa-integral}
 X= \int \left[{\bf c}\!\cdot\!{\bf  j}-{({\bf c}\partial{\bf
       c})\over8\pi}\right] \,d\Omega %
  +\text{surface terms},
\end{equation}
but the explicit calculation shows that this expression does not
contain an integration variable needed to fix the total flux of the
field ${\bf c}$.  To set this flux correctly, we need to add the term
\begin{displaymath}
  \displaystyle\lambda_\mu\left(
    {\bf j}-{\nabla\!\times\!{\bf c} \over4\pi}
  \right)_\mu
\end{displaymath}
with integration over the components of the uniform auxiliary field
$\lambda_\mu$.  This is equivalent to restoring the integration over
the uniform part of the field ${\bf c}$ removed by the surface
patching terms; the whole expression still remains gauge-invariant.
With the addition of this constraint, the integration in
Eq.~(\ref{eq:fa-integral}) yields {\em exactly\/} the
self-coupling~(\ref{eq:fa-self-coupling}).

Because it is always an integer multiple of $2\pi$, the
expression~(\ref{eq:fa-integral}) can be introduced as an additional
phase to the partition function~(\ref{eq:def-csxy}) without changing
its value.  As a result, the full partition function contains a
minimal coupling of the integer-valued current ${\bf j}$ to the sum of
two fields ${\bf a}'={\bf a}+{\bf c}$ with their respective gauge
kernels ${\cal K}^{\mu\nu}$ and
${\cal{}K}_0^{\mu\nu}=ie^{\mu\nu\rho}\nabla_\rho/(4\pi)$.  After
integrating away\cite{Pryadko-95} the difference of these fields, the
resulting model has the gauge-invariant coupling to the field ${\bf
  a}'$ with the gauge kernel
\begin{equation}
  \label{eq:fa-new-kernel}
  {\cal K}'=
  \left({\cal K}^{-1}+{\cal K}_0^{-1}\right)^{-1}
\end{equation}
with the Chern-Simons coefficient $\alpha'=\alpha+2$.  Since both
fields ${\bf a}$ and ${\bf c}$ were constrained to have a fixed total
magnetic flux, the combined field ${\bf a}'$ has the magnetic flux
\begin{equation}
  \label{eq:fa-flux}
  {\bf \Phi}'={\bf \Phi}+4\pi{\bf I}
\end{equation}
in every direction.  Clearly, the total magnetic flux increased by
twice the number of the particle's world lines propagating through the
system in the same direction, while the total particle current ${\bf
  I}$ remains unchanged.

Similarly to the duality, the flux attachment transformation preserves
the mean-field constraint~(\ref{eq:3d-mf-constraint}) for the models
with purely Chern-Simons gauge kernel~(\ref{eq:def-purely-cs}).
Therefore, imposing such a constraint once for the initial model
immediately sets it for the whole hierarchy of models generated by
the duality and flux attachment transformations, in agreement with our
expectations for the models describing plateau transitions in quantum
Hall systems.

\section{Correspondence Laws}
\label{sec:correspondence}

The hierarchy transformations of the CS \xy\ model change the
microscopic dimensionless charge $\alpha$ defined through the gauge
coupling kernel~(\ref{eq:gen-cs-kernel}), and, more importantly, they
also modify the global parameters, the Euclidean current and the
magnetic flux, in a universal fashion.  It turns out that these global
transformations are formally equivalent to the Correspondence Laws,
suggested\cite{Kivelson-92} and confirmed
experimentally\cite{Glozman-95,Shahar-95A,Shahar-95B,Shahar-96A,Shahar-96B},
as the symmetries of the quantum Hall phase diagram.  This analogy can
be seen in the real-time representation, which requires treating the
charge-density $\rho$ and the lateral current $\vec{\hbox{\it\j}}$
separately.  

\paragraph{Symmetries of the phase diagram.} 
The Wick rotation transforms the third component ${\bf j}_z$ of the
Euclidean current into the charge density $\rho$, while
the corresponding component of the total magnetic field $B_{\rm
  tot}=\nabla\!\times\!({\bf a}+{\bf A})$ is transformed to the
component of the magnetic field normal to the two-dimensional plane of
motion of particles.  It is convenient to express the global
transformations of the \xy\ model in terms of the bosonic
filling fraction $\nu_{\rm b}= 2\pi\langle\rho\rangle/\langle B_{\rm
  tot}\rangle$:
\begin{eqnarray}
  \mbox{reflection (\protect\ref{eq:reflection}):}&\qquad&
  \label{eq:nub-reflect}%
  \nu_{\rm b}\rightarrow-\nu_{\rm b},\\ %
  \mbox{duality (\protect\ref{eq:total-dual-current}),
    (\protect\ref{eq:3d-dual-flux}):} &\qquad& 
  \label{eq:nub-duality}%
  \nu_{\rm b}\rightarrow-\nu_{\rm b}^{-1},\\ %
  \mbox{flux attachment (\ref{eq:fa-flux}):}
  &\qquad& \label{eq:nub-fattach}%
  \nu_{\rm b}^{-1}\rightarrow\nu_{\rm b}^{-1}+2,
\end{eqnarray}
which are formally equivalent to the Correspondence
Laws\cite{Kivelson-92} for fermions.  Indeed, by attaching a flux
quantum to every boson, we can convert it to a fermion in a modified
magnetic field, their filling fractions related through the identity
$1/\nu_{\rm b}=1/\nu_{\rm f}-1$.  Then each of the basic
transformations for fermions become equivalent to a combination of
transformations~(\ref{eq:nub-reflect})--(\ref{eq:nub-fattach}), as
explicitly shown in Table~\ref{tab:correspondence}.  

\begin{table}[bhtp]
  \begin{center}
    \leavevmode
    \begin{tabular}[c]{l|l|l}
      Transformation&  Fermions & Bosons\\
      \hline
      \mbox{Landau level addition} &
      $\nu_{\rm f}\rightarrow\nu_{\rm f}+1$&%\;\Leftrightarrow&\;
      $\nu_{\rm b}\rightarrow-\nu_{\rm b}^{-1}-2$ \\
      \mbox{flux attachment}&
      $\nu_{\rm f}^{-1}\rightarrow\nu_{\rm f}^{-1}+2$&%\Leftrightarrow&
      $\nu_{\rm b}^{-1}\rightarrow\nu_{\rm b}^{-1}+2$\\
      \mbox{particle-hole symmetry}&
      $\nu_{\rm f}\rightarrow1-\nu_{\rm f}$&%\Leftrightarrow&
      $\nu_{\rm b}\rightarrow\nu_{\rm b}^{-1}$
    \end{tabular}
  \end{center}
  \caption{The Laws of Correspondence in terms of Fer\-mi\-onic
    $\nu_{\rm f}$ and Bosonic $\nu_{\rm b}$ filling fractions related
    by the identity $1/\nu_{\rm b}=1/\nu_{\rm f}-1$.} 
  \label{tab:correspondence}
\end{table}

Experimentally\cite{Shahar-95B,Shahar-96A,Shahar-96B}, the duality
symmetry was observed for the transitions between the quantum Hall
liquid and insulator.  Specifically, the 2DEG demonstrates reciprocal
longitudinal resistance and the mirror symmetry of non-linear $I$--$V$
curves in the points at the phase diagram, related by the expression
\begin{equation}
  \label{eq:shahar-duality}
  \left(\alpha-{\nu}^{-1}\right)\left(\alpha-{\tilde\nu}^{-1}\right)=1, 
\end{equation}
where $\alpha$ is an odd integer value of the quantized Hall
resistance in the corresponding plateaus, measured in the units of
$h/e^2$, and $\nu=2\pi\rho/B$ is the measured filling fraction of the
electrons, which is varied across the transition.  In terms of the
``bosonic'' filling fractions $\nu_{\rm b}=(\alpha-{\nu}^{-1})^{-1}$,
this relationship can be written simply as $\nu_{\rm
  b}\,\tilde\nu_{\rm b}=1$, which is equivalent to the combination of
the duality~(\ref{eq:nub-duality}) with the
reflection~(\ref{eq:nub-reflect}).  Note, that the measured value of
the filling fraction $\nu_c$ in the transition point is close but not
exactly equal to the theoretical value $\nu^\ast_c=(\alpha+1)^{-1}$,
implying that the quasiparticles are not in one-to-one correspondence
with the original electrons.  A partial mixing of the opposite spin
states, not included in the theoretical model, provides a qualitative
explanation for this discrepancy.

\paragraph{Symmetries of  transport coefficients.}

The remaining two components of the Euclidean current ${\bf
  j}=(\rho,i\vec{\hbox{\it\j}})$ originate from the lateral current
$\vec{\hbox{\it\j}}$ of the $2\!+\!1$ dimensional system, while the in-plane
Euclidean magnetic field is generated by the electric
field
\begin{equation}
  \label{eq:euclidean-b-to-e}
  {{\bf B}_x}\rightarrow  i{E_y},\quad   
  {{\bf B}_y}\rightarrow -i{E_x}. 
\end{equation}
The rotation by $\pi/2$ is needed to transform the components of the
axial vector ${\bf B}$ into the polar vector ${\vec E}$; the inversion
of the handedness changes both the sign of ${\bf B}$ and the sign of
the rotation angle, so that the relation~(\ref{eq:euclidean-b-to-e})
remains valid.  Unlike that for the usual charge density and magnetic
field, the Wick rotation of spatial components involve the complex
phase $i=e^{i\pi/2}$, indicating that the analytic continuation in the
values of these fields is needed.  Any singularity encountered during
this procedure would invalidate the current--voltage mapping.  This is
certainly related to the fact that the presence of the electric field
$\vec E$ and the current density $\vec{\hbox{\it\j}}$ may involve
dissipation.  However, this mapping is valid in all orders of the
perturbation expansion in powers of the electric field $\vec{E}$; the
actual region of validity of this procedure is determined by the
convergence of the perturbation series, as well as the presence and
the relative importance of non-perturbative terms.
%%Clearly, choosing a ``wrong'' value of
%%$\alpha$ for a given physical system (specified by the disorder,
%%charge density and the magnetic field) must invalidate the
%%perturbation theory along with the incorrect value of the critical
%%resistance.

The duality interchanges the components of the current density with
the appropriate components of the rotated electric field.  In terms of
physical electric field and current density, the duality relationship
can be written as
\begin{equation}
  \label{eq:dual-current}
  \left({\tilde
      E_{x,y}\over2\pi\,\tilde{\hbox{\it\j}}_{y,x}}\pm\alpha\right) 
  \left({E_{x,y}\over2\pi \,\hbox{{\it j\/}}_{y,x}}\pm\alpha\right)
  =1, 
\end{equation}
where the first (second) indices correspond to top (bottom) signs.  In
the linear regime the substitution
${\vec E}=\hat\rho\,\vec{\hbox{\it\j}}$ and %
$\vec{\tilde E}=\hat{\tilde\rho\,}\vec{\tilde{\hbox{\it\j}}}$ shows
that the equation~(\ref{eq:dual-current}) is equivalent to the
inversion of the bosonic resistivity tensor
\begin{equation}
  \label{eq:dual-sigma}
  \hat\rho_b\equiv{{\hat{\sigma}}_b}^{-1}=\hat{\tilde\sigma}_b
  ={\hat{\tilde\rho\,}\!}_b^{-1}
\end{equation}
between the mutually dual points.  Here the bosonic resistivity
tensors are related to the usual resistivity tensors, measured in the
units of $hc/e=2\pi\,\hbar c /e$, by the relationship
\begin{equation}
  \label{eq:attached-resistivity}
  \hat\rho_b=\hat\rho-\alpha\left(
  \begin{array}[c]{cc}
    0&1\\-1&0
  \end{array}\right).
\end{equation}
As pointed out in Ref.\CITE{Shimshoni-96}, Eq.~(\ref{eq:dual-sigma})
corresponds to the inversion of the longitudinal resistivity
$\tilde\rho_{xx}=1/\rho_{xx}$ observed in experiment\cite{Shahar-95B}
as long as the Hall resistivity remains equal to its quantized value
$\rho_{xy}=\alpha$, or, equivalently,
\begin{displaymath}
  \sigma_{xy}^2+\sigma_{xx}^2=\alpha^{-1}\sigma_{xy}.
\end{displaymath}
This statement, known as the Semicircle Law, is associated with the
presence of disorder, and is known to hold exactly for two-component
systems\cite{Dykhne-94,Ruzin-95}.  

\paragraph{Universality of phase transitions.}

Perturbati\-ve\-ly\cite{Wen-Wu-93,Wei-93,Pryadko-94}, the universality
of phase transitions in Chern-Simons models does not seem to be a
probable outcome.  However, the universality was
proven\cite{Pryadko-95} for the transitions in the clean CS \xy\ model
where the ``relativistic'' symmetry is not broken by external magnetic
field or total charge density.  In the absence of these fields the
independent parameter $\nu$ is absent, and the hierarchy mapping
between the vicinities of phase transitions at different values of
$\alpha$ is complete.  The exact universality class depends on whether
the given fractional value of the Chern-Simons coefficient originates
from bosons or fermions, but it is not likely to correspond to that of
any phase transition experimentally observed in quantum Hall systems.

In the extended CS \xy\ model~(\ref{eq:def-csxy}), where both the
external magnetic field and the charge density can be present, there
are at least two independent parameters that characterize the state of
the equilibrium system: the filling fraction $\nu$ and the
Chern-Simons coefficient $\alpha$.  Transforming both of them at each
step, it is impossible to construct a mapping between two quantum Hall
states of electrons ($\alpha=1$).  Even though the proof of the
universality\cite{Pryadko-95} remains formally valid, in the presence
of non-zero filling fraction the universality applies not to the
original electrons, but to vortices, particles with the transformed
Chern-Simons coefficient, at the dual filling fraction.  In the
original formulation of the hierarchy theory\cite{Kivelson-92}, this
problem is resolved by making an {\em assumption\/} about the
independence of the bosonic current-current correlation function of
the Chern-Simons coefficient.  So far, even though an ample
experimental evidence confirming this non-trivial assumption is
available, a rigorous theoretical proof, or even an effective model
that would display this behavior, is missing.

Any possible scenario explaining this puzzle must include the
disorder.  For example, if the disorder-associated scattering turns
out strong enough to screen the correlations associated with the
Chern-Simons interaction at large distances, the effective
disorder-averaged model will be independent of the Chern-Simons
coefficient.  Since both the duality and the flux attachment preserve
the shape of the disorder potential, it is conceivable that the
disorder-averaged model would nevertheless inherit the original
hierarchy symmetries, at least in some range of parameters.  This
point of view is supported by the bulk of
experimental data\cite{Wei-88,Engel-90,Koch-91A,Glozman-95,%
  Shahar-95A,Shahar-95B,Shahar-96A,Shahar-96B}, which indicates that
the fractional and integer quantum Hall transitions are in the same
universality class, and their vicinities are at least approximately
related by the Correspondence Laws.

\section*{Conclusion}

The Chern-Simons \xy\ model with external fields considered here was
constructed as a lattice regularization of the CSLG model.  Even
though this model is written in rotationally (relativistically) {\em
  covariant\/} form~(\ref{eq:def-csxy})--(\ref{eq:def-scalar}), the
non-zero charge-current density and external magnetic and electric
fields break the Lorentz symmetry in exactly the same way as it is
broken in the original continuum model.  In addition, the studied
model demonstrates the same non-local symmetries as the CSLG
model---only at the lattice these symmetries can be formulated as
exact statements, without the limitations of the RPA approximation
used in the continuum model.

Unfortunately, neither model clears the issue of the observed
universality of phase transition.  Note, however, that both the
duality and the flux attachment transformation remain exact in the
Chern-Simons \xy\ model even in the presence of disorder; the latter
is merely carried from one representation to another without any
significant change of the statistical properties.  It is easy to
check\cite{Pryadko-97-replicas} that in the replicated version of this
model the disorder averaging commutes with the duality transformation.
This gives a hope that there exists an effective description of
diffusion in this strongly-interacting system, independent of the
Chern-Simons coupling and, therefore, symmetric with respect to
hierarchy transformations.

Apart from this issue of universality, the analyzed CS \xy\ model
appears to be a perfect candidate for providing an effective
description of the quantum Hall phase transitions.  In this
second-quantized model the non-local duality and flux attachment
transformations can be performed exactly.  These transformations are
expressed in terms of system-averaged charge and current density and
also the electric and magnetic fields; they generate the
experimentally observed global symmetries of the phase diagram known
as the Correspondence Laws\cite{Kivelson-92}, and the
nonlinear current--voltage mapping between mutually dual
points\cite{Shahar-95B}.  The global nature of the transformed
parameters imply that they are not subject to renormalization.
Moreover, the derived duality relationship is so robust and
insensitive to particular details of both interaction and disorder,
that the immediately following $I$--$V$ reflection symmetry could be
viewed as a signature of the duality of this kind.

In this context, a sim\-i\-lar and so far un\-ex\-plained
current--voltage symmetry observed\cite{Simonian-96,Kravchenko-96} in
the con\-duc\-tor--insulator transition in silicon MOSFETS at zero
magnetic field appears especially suggesting.  Certainly, there
remains a more mundane possibility that the observed symmetry is a
mere consequence of the universal critical scaling.
%%Even then, the
%%unusual size of the critical region would require at least an explanation.

\section*{Acknowledgments}
I am grateful to Robert~B.~Laughlin for pointing out that the magnetic
fields require special attention in finite systems---the idea which
eventually lead to this paper, to Shou-Cheng~Zhang for guiding me into
this field, and especially to Steve Kivelson for many inspiring
discussions.

\appendix
\section{Hermiticity of the Chern-Simons coupling}
\label{appen:hermiticity}

Here I derive the conditions of Hermiticity 
\begin{eqnarray}
  \label{eq:cs-hermiticity}
  \int (a\partial b) d\Omega&\!-\!&
  \sum_i\int_{S_i} G_i \nabla\times{\bf b}\cdot
  d{\bf s}\\&=&
  \int (b\partial a) d\Omega-\sum_i\int_{S_i} F_i \nabla\times{\bf
  a}\cdot d{\bf s}\nonumber
\end{eqnarray}
for the gauge-invariant Chern-Simons coupling in finite systems, with
the patching functions $G_i$ and $F_i$ defined (up to the additive
constants specified below) by the expressions
\begin{equation}
  \label{eq:ap-patch-func}
  \nabla G_i=\Incr_i{\bf a},\qquad\nabla F_i=\Incr_i{\bf b}
\end{equation}
Consider the difference of two volume integrals
\begin{eqnarray*}
  J=\int{\bf a}\cdot\nabla\times{\bf b}\;d\Omega&
  -&\int{\bf b}\cdot\nabla\times{\bf a}\;d\Omega\\
  = \int\nabla\cdot({\bf b}\times{\bf a})\;d\Omega
  &=&\oint{\bf b}\times{\bf a}\cdot d{\bf s}
%  +\sum_{\bf L}  F^a_{\bf L} \nabla\cdot{\bf b} \cdot d{\bf A}
\end{eqnarray*}
transformed to the integral over the full outer surface of the system,
which can be split into the contributions from three pairs of parallel
faces,
\begin{eqnarray}
  J&=&\sum_{i=\{x,y,z\}}J_i,\label{eq:int-split}
  \nonumber \\
  J_i&=& %
  \int_{S_i}\nabla F_i\times\left({\bf a}-{1\over2}\nabla G_i\right)
  \cdot d{\bf s}
  \\ &\relax& %
  - \int_{S_i}\nabla G_i\times\left({\bf b}-{1\over2}\nabla F_i\right)
  \cdot d{\bf s}. \nonumber
\end{eqnarray}
In each term the integration is limited to the front $S_x$, right
$S_y$ or top $S_z$ sides of the system as shown in
FIG.~\ref{fig:cube}, and the patching
functions~(\ref{eq:ap-patch-func}) were used to simplify the
increments of the gauge fields across the system.  Each of these six
integrals can be further transformed using the vector identity
$$
\nabla F\times{\bf a}
=\nabla \times \left(  {\bf a}\, F\right)- F\, \nabla\times{\bf a}
$$
into the sum of the contour integral and the surface integral
\begin{displaymath}
  \int\limits_{S}\!\nabla  
  F\times\left(\!\!{\bf a}-{\nabla G\over2}\!\right)
  d{\bf s} = %
  \!\oint\limits_C\! 
  F \left(\!\!{\bf a}-{\nabla G\over2}\!\right)\!\cdot d {\bf l} %
  -\!\int\limits_S\! F\nabla\!\times\! {\bf a}\cdot d{\bf s}.
\end{displaymath}
In every case the surface integral cancels with one of the surface
patching terms in $J$.  Therefore, the Chern-Simons coupling with
appropriate edge patching terms is a Hermitian operator as specified
by Eq.~(\ref{eq:cs-hermiticity}) {\em if and only if\/} the sum of all
line integrals
\begin{equation}
  \label{eq:line-integrals}
  L=\sum_i\oint_{C_i} F_i\left({\bf a}-{1\over2}G_i\right)\cdot
  d{\bf l}
  - \biggl[
  \begin{array}[c]{c}
    a\leftrightarrow b\\
    G\leftrightarrow F
  \end{array}
  \biggr]
\end{equation}
vanishes.

The closed line integrals in Eq.~(\ref{eq:line-integrals}) can be
simplified by grouping together the contributions of the parallel
parts of each contour, and once again using the patching functions for
the differences of the gauge fields
\begin{eqnarray}
  \label{eq-line-integrals-one}
 L   &=& e^{xyz}\int_0^{{\bf L}_y} dy \biggl\{
  F_z\left(a_y-{1\over2}G_{z,y}\right)
  \nonumber \\ %
  &\relax&-\left(F_z-\Incr_x F_z\right)
  \left(a_y-{1\over2}G_{z,y}-G_{x,y}+{1\over2}\Incr_x G_{z,y}\right)
  \biggr\}
  \nonumber\\
  &\relax&- \biggl[
  \begin{array}[c]{c}
    a\leftrightarrow b\\
    G\leftrightarrow F
  \end{array}
  \biggr].
\end{eqnarray}
Here we imply the summation over all permutations of the coordinates,
$G_{z,y}\equiv\partial_y G_z$, and the functions are evaluated at the
segment $({\bf L}_x,y,{\bf L}_z)$ or its equivalents generated by cyclic
permutations of the coordinates.  It is convenient to use the
symmetries of the expression~(\ref{eq-line-integrals-one}) and write
symbolically
\begin{eqnarray}
  \label{eq:line-integrals-two}
  L&=&\sum_{12} \int_0^{{\bf L}_y} \biggl\{
  \left(F_z-\Incr_x F_z\right) \left(G_{x,y}-{1\over2}\Incr_x
  G_{z,y}\right)
  \nonumber\\
  &\relax& + \Incr_x F_z
  \left(a_y-{1\over2}G_{z,y}-G_{x,y}\right)\biggr\}dy,
\end{eqnarray}
implying the summation over six permutations of the coordinates and
two additional permutations of the gauge fields $a$ and $b$ with their
corresponding parity factors.  This way one does not need to write all
terms explicitly in order to simplify the
expression~(\ref{eq:line-integrals-two}).  For example, the integral
\begin{eqnarray*}
  \sum_{12}\int F_z G_{x,y} \,dy
  &=&{1\over2}\sum_{12}\int\left(F_zG_{x,y}+G_xF_{z,y}\right)\,dy\\ %
  &=&{1\over2}\sum_{12}\left.(F_z G_x)\right|_0^{{\bf L}_y}\equiv
  {1\over2}\sum_{12}\Incr_y (F_z G_x)\\ %
  &\hskip-1cm=\hskip1cm&\hskip-1cm
  {1\over2}\sum_{12}\Incr_y F_z\,G_x+F_z\,\Incr_y G_x-
  \Incr_yF_z\,\Incr_y G_x. 
\end{eqnarray*}
Some additional simplification can be achieved by using the expression
for the total flux of the gauge field
$$ %
\Incr_y F_x-\Incr_x F_y=\oint_{C_z} {\bf a}\cdot d{\bf l}={\bf\Phi}_z^a
$$ %
in $z$-direction, and similar expressions for other components of
${\bf a}$ and the field ${\bf b}$.  After some rather tedious but
straightforward algebra the expression~(\ref{eq:line-integrals-two})
simplifies to
\begin{equation}
  \label{eq:line-integrals-ans}
  L=\sum_6 {\bf\Phi}_y^a\left(F_y-\int_0^{{\bf L}_y} b_y dy
    +{1\over2}{\bf\Phi}_z^b-{1\over2}{\bf\Phi}_x^b\right).
\end{equation}
The function $F_y$ in this expression is evaluated at the vertex
$({\bf L}_x,{\bf L}_y,{\bf L}_z)$ and the integration in $y$ is performed along the line
$x={\bf L}_x$, $z={\bf L}_z$.  

With the help of Eq.~(\ref{eq:line-integrals-ans}) 
we can fix the so far arbitrary constants in the
definition~(\ref{eq:ap-patch-func}) of the surface
patching functions $F_i$ and $G_i$.  Specifically, by defining
\begin{equation}
  \label{eq:gauge-patch-const}
  F_y({\bf L}_x,{\bf L}_y,{\bf L}_z)=\int\limits_0^{{\bf L}_y} b_y({\bf L}_x,y,{\bf L}_z) dy
    -{1\over2}{\bf\Phi}_z^b+{1\over2}{\bf\Phi}_x^b
\end{equation}
(and similar definitions for other functions $F_i$ and $G_i$) we can
completely suppress the contour integrals~$L$
                                %(\ref{eq:line-integrals})   
and therefore render the Chern-Simons coupling with the surface term a
Hermitian operator as defined by Eq.~(\ref{eq:cs-hermiticity}).  It is
important that the condition~(\ref{eq:gauge-patch-const}) is not
modified by the gauge transformation~(\ref{gauge-transf}) because the
fluxes ${\bf\Phi}_\mu$ are gauge-invariant.  Therefore, both the
Hermiticity and gauge-invariance can hold at the same time.

%\bibliography{main,add,more}
%\bibliographystyle{prsty}

\end{document}